\documentclass[aps,pra,reprint]{revtex4-2}
\usepackage{amsmath,amssymb}
\usepackage{graphicx}
\usepackage{bm}
\usepackage{stmaryrd}   
\usepackage{seqsplit}   
\usepackage{hyperref}   

\providecommand{\enquote}[1]{``#1''}

\begin{document}

\title{Ultrasensitive measurement of phase dynamics by time-holographic
dark-port readout of a marginally imbalanced interferometer}

\author{F. Silva}
\author{A. Molina}
\author{A. Esteban-Mart\'{i}n}
\author{J. Garc\'{i}a-Monreal}
\author{G. J. de Valc\'{a}rcel}
\email{german.valcarcel@uv.es}
\affiliation{Universitat de Val\`{e}ncia, Av. Vicent Andr\'{e}s Estell\'{e}s, 19.
46100 Burjassot, Val\`{e}ncia, Spain}

\date{\today}

\begin{abstract}
Resolving displacements and surface irregularities far below a nanometer
is now indispensable across fields, from precision manufacturing and
biotechnology to nanoscience and fundamental metrology. High-precision
optical interferometry reaches this $\lambda/1{,}000$ regime by reading
out optical power, which provides a smooth response over the measurement
range. Here we instead read out the complex dark-port field of a
marginally imbalanced interferometer, whose phase response to small
internal phase deviations $\phi_{\textrm{in}}$ near destructive
interference becomes extremely steep. For a small arm-amplitude
imbalance $\mu$, the tangent $T_{\textrm{out}}$ of the dark-port phase
reduces to $\mu^{-1}\phi_{\textrm{in}}$, defining a phase magnification
$M=\mu^{-1}$. Crucially, this tangent estimator stays linear over a
range $|\phi_{\textrm{in}}|\lesssim\sqrt{|\mu|}$---much larger than the
imbalance $|\mu|$ itself for $|\mu|\ll1$---so that fine temporal
substructure is resolved within much larger phase excursions.
$T_{\textrm{out}}$ is synthesized from the dark-port complex field by
optical time-domain holography and Fourier demodulation---a heterodyne
readout that is intrinsically immune to laser relative-intensity noise
at any magnification, and gives time-resolved access to phase dynamics.
In a proof-of-principle Mach--Zehnder experiment, without optimization
or environmental isolation, we demonstrate $|M|>10^3$ and optical-path
resolution better than $\lambda/100{,}000$ ($\approx5~\textrm{pm}$),
with a shot-noise floor orders of magnitude below this
environment-limited value. The approach is compact, tunable, and
compatible with existing interferometric platforms, offering a practical
route to enhanced phase sensitivity in metrology, sensing, and optical
instrumentation.
\end{abstract}

\maketitle

\section{Introduction}

Optical interferometry~\cite{Hariharan-book} ranks among the most precise techniques for measuring departures from a reference state. Its performance stems from the relative-phase information encoded in the superposition of mutually coherent optical waves, enabling the detection of optical-path changes far below the optical wavelength. Instruments based on this principle---most notably high-precision Michelson~\cite{Michelson} and Mach--Zehnder~\cite{Zehnder,Mach} interferometers---achieve sub-nanometer resolution through output-power readout. This level of performance has become indispensable across industrial, technological, and fundamental-science applications, including precision manufacturing and testing, metrology, and sensing.

Power readout provides a smooth response and is the conventional mode of interferometric operation. The maximum sensitivity occurs at phase quadrature, where the phase difference between the interfering waves is $\pi/2$ and the output power varies linearly with small internal phase changes around this point. At the dark fringe, where the phase difference is $\pi$, the situation is reversed for the two most natural observables. The power sensitivity is minimal, since the output power varies quadratically with phase variations. The phase of the output field, however, becomes extremely sensitive to internal phase changes near destructive interference~\cite{Hecht-book}. This complementary dark-port phase response, inaccessible to conventional power readout, suggests an alternative readout channel for phase metrology.

Several strategies have been developed to extend interferometric resolution beyond the direct power response of a simple two-beam interferometer. Multi-pass~\cite{F-P,GTI} and double-pass~\cite{Harimoto-2023} arrangements increase the accumulated optical phase by making the perturbation act more than once on the probing field, and have been used, for example, in nanoscale dimensional metrology~\cite{Ren-2008}. Resonant or feedback-assisted interferometers provide another route, using optical cavities or multiport feedback to reshape the phase response and enhance the slope around selected working points~\cite{Schwarze-2023}. Gravitational-wave (GW) detectors such as LIGO and Virgo represent the pinnacle of optical interferometry, detecting displacements at the attometer scale~\cite{GW} through the rigorous suppression of both technical and fundamental noise sources. In this case, multi-pass operation near the dark port is combined with DC-readout schemes~\cite{Hild-2009,Fricke-2012}, further improved by the deployment of squeezed light to surpass the standard quantum limit~\cite{Jia-2024}. More fundamental approaches have also examined the role of asymmetry in interferometric readout~\cite{Ariel-2015,Ariel-2018,Ariel-2020}, showing that enhanced sensitivity can arise not only from noise reduction, but also from engineering the optical response and the readout channel.

Here we follow this latter route and analyze phase readout at the dark port of a marginally imbalanced interferometer. The imbalance prevents complete cancellation at destructive interference, while preserving the strong phase response of the output field. We first revisit imperfect destructive interference between two waves with slightly different intensities. We characterize the extreme sensitivity of the tangent of the superposition phase to tiny deviations of the relative phase from $\pi$, providing a magnified estimator whose gain is controlled by the intensity asymmetry. This result is directly applicable to single-pass interferometers such as Mach--Zehnder and Michelson interferometers.

We then report a proof-of-principle Mach--Zehnder experiment in which time-domain holography gives direct access to the dark-port complex field. Magnifications above $1{,}000\,\times$ enable the ultrasensitive measurement of tiny temporal phase variations, corresponding to optical-path changes in the picometer range. Alongside the experiment, we provide estimates of the fundamental and technical noise limits of the readout---the shot-noise floor and the laser and detection channels---which set the resolution ultimately attainable and guide the choice of components for further improvement.

\section{Phase magnification near two-wave imperfect destructive interference}

A two-beam optical interferometer, such as a Michelson or Mach--Zehnder interferometer, provides the coherent superposition of two light waves with equal angular frequency $\omega$. Assuming plane waves, we denote their individual complex electric fields at the detection plane by $E_j(t)=\kappa A_j\textrm{e}^{\textrm{i}(\omega t+\phi_j)}$, where $j=1,2$ labels the wave, $\kappa\equiv\sqrt{{2\hslash\omega}/{\varepsilon_0 c S}}$ is a normalization constant ($S$ is the effective beam cross-sectional area at the detector plane), chosen so that $I_j\equiv A_j^2=|E_j|^2/\kappa^2$ has the meaning of photon flux onto the photodetector (number of photons per unit time); equivalently, $P_j\equiv\hslash\omega I_j$ is the optical power impinging the photodetector. We will interchangeably refer to $I_j$ as photon flux or as intensity throughout. We introduce the differential phase $\phi_{\textrm{in}}\equiv\phi_2-\phi_1-\pi$ mod $2\pi$, which measures the internal deviation from the condition of maximum destructive interference, corresponding to $\phi_{\textrm{in}}=0$, as we focus on a dark, or antisymmetric, port. Taking $\phi_1=0$ as the phase reference, we have $\phi_2=\pi+\phi_{\textrm{in}}$. The total output field is then $E_{\textrm{out}}=E_1+E_2\equiv \kappa A_{\textrm{out}}\textrm{e}^{\textrm{i}(\omega t+\phi_{\textrm{out}})}$, where $A_{\textrm{out}}$ and $\phi_{\textrm{out}}$ satisfy the well-known relations~\cite{Hecht-book}
\begin{subequations}\label{eqs:Hecht}
\begin{gather}
    \tan\phi_{\textrm{out}}=\frac{A_2\sin\phi_{\textrm{in}}}{A_2\cos\phi_{\textrm{in}}-A_1},\label{phi-Hecht}\\
    I_{\textrm{out}}=I_1+I_2-2\sqrt{I_1I_2}\cos\phi_{\textrm{in}},\label{I-Hecht}
\end{gather}
\end{subequations}
with $I_{\textrm{out}}\equiv A_{\textrm{out}}^2$. Since all fields are referred to the detection plane, whenever we speak of arm or internal observables, such as fields or powers, we mean their values at that plane.

In the equal-amplitude, or balanced, case $P_1=P_2$, the minimum power at the dark port, $P_{\textrm{out}}^{\textrm{min}}$, vanishes, yielding maximum fringe contrast. In practice, however, some degree of imbalance is always present, either intentionally or incidentally. We quantify it through a dimensionless parameter $\mu$, defined through $P_1=(1-\mu)^2P_2$ or,
\begin{equation}\label{def:mu}
\mu\equiv 1-{A_1}/{A_2},\qquad \vert\mu\vert\ll 1,
\end{equation}
which will be assumed small throughout. In terms of $\mu$, the phase transfer function (\ref{phi-Hecht}) can be written as
\begin{equation}\label{def:T_mu}
\tan\phi_{\textrm{out}}=\frac{\sin\phi_{\textrm{in}}}{\mu+\cos\phi_{\textrm{in}}-1}\equiv T_\mu(\phi_{\textrm{in}}),
\end{equation}
which defines the function $T_\mu$. In the perfectly balanced case ($\mu=0$) Eq.~(\ref{def:T_mu}) can be equivalently written as
$\phi_{\textrm{out}}=\phi_{\textrm{in}}/2+\operatorname{sign}(\phi_{\textrm{in}})\pi/2$. This relation is piecewise linear, with slope $1/2$ and a $\pi$ discontinuity at $\phi_{\textrm{in}}=0$, enabling highly sensitive detection of zero-phase crossings \cite{OTD}. A finite imbalance $\mu\ne0$ turns this $\pi$ discontinuity into a steep but continuous transition.

Near the dark fringe ($|\phi_{\textrm{in}}|\ll1$) Eq.~(\ref{def:T_mu}) reduces to
\begin{equation}\label{lin-tan}
\tan\phi_{\textrm{out}}\simeq\mu^{-1}\phi_{\textrm{in}},\qquad\text{for }\vert\phi_{\textrm{in}}\vert\lesssim0.3\vert\mu\vert^{1/2},
\end{equation}
showing a phase magnification $M\equiv\mu^{-1}$, which increases as the imbalance is reduced. Equation~(\ref{lin-tan}) has an $\mathcal{O}(\phi_{\textrm{in}}^3)$ correction, whose departure from linearity can be quantified through the relative error $\delta_\mu(\phi_{\textrm{in}})\equiv\left\vert(\mu\tan\phi_{\textrm{out}}-\phi_{\textrm{in}})/\phi_{\textrm{in}}\right\vert$. Using Eq.~(\ref{def:T_mu}) and assuming both $\mu$ and $\phi_{\textrm{in}}$ to be small gives, to leading order,
$\delta_\mu(\phi_{\textrm{in}})=\phi_{\textrm{in}}^2/(2\vert\mu\vert)$. Therefore, requiring $\delta_\mu(\phi_{\textrm{in}})<5\%$ gives $\vert\phi_{\textrm{in}}\vert\lesssim0.3\sqrt{\vert\mu\vert}$, which justifies the linear range stated in Eq.~(\ref{lin-tan}). If a 10\% relative error is acceptable, the effective linear range extends to $\vert\phi_{\textrm{in}}\vert\lesssim0.45\sqrt{\vert\mu\vert}$.

Relation~(\ref{lin-tan}) only requires $\phi_{\textrm{in}}$ to be sufficiently small. If, in addition, $\phi_{\textrm{out}}$ is also small, the output phase itself may be approximated as
\begin{equation}\label{lin-Ariel}
\phi_{\textrm{out}}\simeq \mu^{-1}\phi_{\textrm{in}},\qquad\text{for }\vert\phi_{\textrm{in}}\vert\lesssim0.4\vert\mu\vert\,,
\end{equation}
where the stated linear range again corresponds to a relative error below 5\%. For a 10\% tolerance, the condition becomes $|\phi_{\textrm{in}}|\lesssim0.55\vert\mu\vert$. In this case, the relevant relative error,
$\left\vert(\mu\phi_{\textrm{out}}-\phi_{\textrm{in}})/\phi_{\textrm{in}}\right\vert$, behaves as
$\phi_{\textrm{in}}^2/(3\mu^2)$. Up to simple changes of notation, Eq.~(\ref{lin-Ariel}) is the relation used in \cite{Ariel-2018,Ariel-2020} to describe phase-shift-amplified interferometry, see Eq.~(5b) of \cite{Ariel-2018}, with amplification factors reaching $M=35$. Much larger magnifications have been reported in RF interferometry~\cite{Ariel-2015}, where sub-mrad phase variations arise from micron-scale displacements.

\begin{figure}[!ht]
\centering\includegraphics[width=0.7\linewidth]{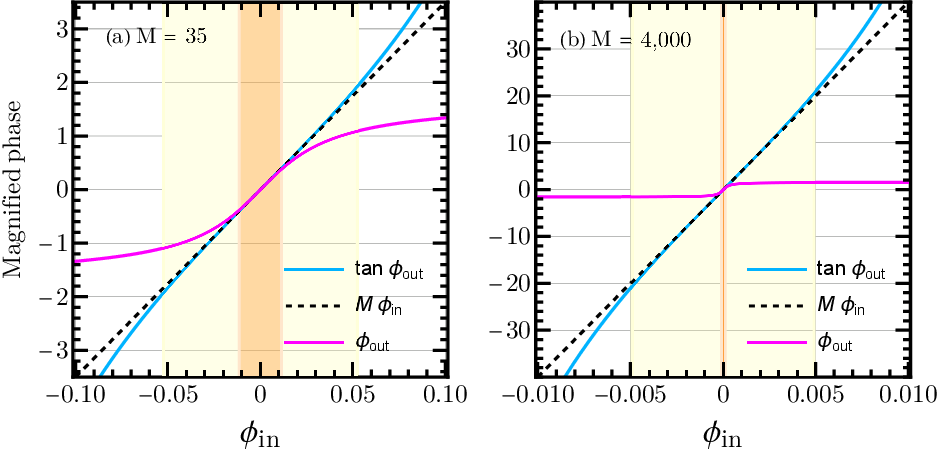}
\caption{Theoretical phase magnification for $M=35$ (a) and $M=4{,}000$ (b). The horizontal axes show the internal phase deviation $\phi_{\textrm{in}}$ from the minimum-output state, at $\phi_{\textrm{in}}=0$; the vertical axes show the magnified output. In each panel, the dashed diagonal is the ideal undistorted magnified signal $M\phi_{\textrm{in}}$; the blue curve, closer to the diagonal, is the tangent readout $\tan\phi_{\textrm{out}}=T_\mu(\phi_{\textrm{in}})$, and the magenta, sigmoid-like curve is the output phase $\phi_{\textrm{out}}=\arctan T_\mu(\phi_{\textrm{in}})$; see Eq.~(\ref{def:T_mu}). The shaded rectangles mark the regions of relative error below $5\%$: the wider yellow ones for the tangent estimator [Eq.~(\ref{lin-tan})], the narrower orange ones for the phase estimator [Eq.~(\ref{lin-Ariel})]. Note the very different horizontal and vertical scales, a direct expression of the magnification.}\label{fig:sigmoid-theory}
\end{figure}

Using Eq.~(\ref{lin-tan}) instead of Eq.~(\ref{lin-Ariel}) enlarges the linearity range by essentially $\vert\mu\vert^{-1/2}\gg1$ in the limit $\vert\mu\vert\ll1$ considered here. This is illustrated in Fig.~\ref{fig:sigmoid-theory} for two magnification factors: (a) $M=35$ ($\mu=1/35$), corresponding to the case reported in \cite{Ariel-2020}, and (b) $M=4{,}000$ ($\mu=1/4{,}000$), compatible with our observations. Importantly, this increase in linear dynamic range is obtained at no additional experimental cost. This point is especially relevant because it allows much smaller internal phase variations to be accessed. Consider first using $\phi_{\textrm{out}}$ as an estimator of $\phi_{\textrm{in}}$ \cite{Ariel-2018,Ariel-2020} within a fixed tolerance, say 5\% as in Eq.~(\ref{lin-Ariel}). For an input signal spanning the range $|\phi_{\textrm{in}}|\leq\phi_{\textrm{in}}^{\textrm{max}}$, the largest usable magnification $M_{\textrm{max}}$ must satisfy
$M_{\textrm{max}}\phi_{\textrm{in}}^{\textrm{max}}=0.4$. The corresponding output phase is then
$\phi_{\textrm{out}}=M_{\textrm{max}}\phi_{\textrm{in}}=0.4(\phi_{\textrm{in}}/\phi_{\textrm{in}}^{\textrm{max}})$, which may be too small to resolve substructure occurring on scales much smaller than $\phi_{\textrm{in}}^{\textrm{max}}$. By contrast, using $\tan\phi_{\textrm{out}}$ as the estimator within the same tolerance, see Eq.~(\ref{lin-tan}), gives a maximum admissible magnification
$M_{\textrm{max}}\simeq0.1/(\phi_{\textrm{in}}^{\textrm{max}})^2$. Hence
$\tan\phi_{\textrm{out}}\simeq(0.1/\phi_{\textrm{in}}^{\textrm{max}})(\phi_{\textrm{in}}/\phi_{\textrm{in}}^{\textrm{max}})$, allowing access to much finer phase variations whenever $\phi_{\textrm{in}}^{\textrm{max}}\ll0.1$, which does not represent a practical limitation in the regime of interest. 

Therefore, we propose using $\tan\phi_{\textrm{out}}$ as an estimator of $\phi_{\textrm{in}}$, the two being related by Eq.~(\ref{lin-tan}). This implies that $M=\partial T_{\textrm{out}}/\partial\phi_{\textrm{in}}$ is simply the sensitivity in metrological terms~\cite{VIM}.

\section{Experimental setup: Holographic phase readout}

Leveraging the rationale discussed above requires measuring the imbalance $\mu$ and the tangent $T_{\textrm{out}}\equiv\tan\phi_{\textrm{out}}$ of the dark-port field phase---theoretically given by $T_\mu$ in Eq.~(\ref{def:T_mu}). The latter in its turn requires accessing the output complex field $E_{\textrm{out}}$ itself rather than inferring phase information from the usual interferometric power readout. This must be done with respect to an optical reference. Several reference-based strategies can be used for accurate phase retrieval, including phase-shifting \cite{Carre-1966} and phase-quadrature \cite{Heydemann-1981} methods, as well as the classical synchronous heterodyne detection technique \cite{Ariel-2020}. Here we use the closely related approach of optical time-domain holography \cite{Fernandez-2013}, which is especially well suited for tracking continuous phase dynamics. In direct analogy with the spatial complex-field reconstruction provided by conventional off-axis holography, and following the general space--time imaging duality \cite{Kolner-1994}, optical time-domain holography gives access to the temporal evolution of the complex optical field.

\subsection{The setup and the temporal hologram}

\begin{figure}[!ht]
\centering\includegraphics[width=0.7\linewidth]{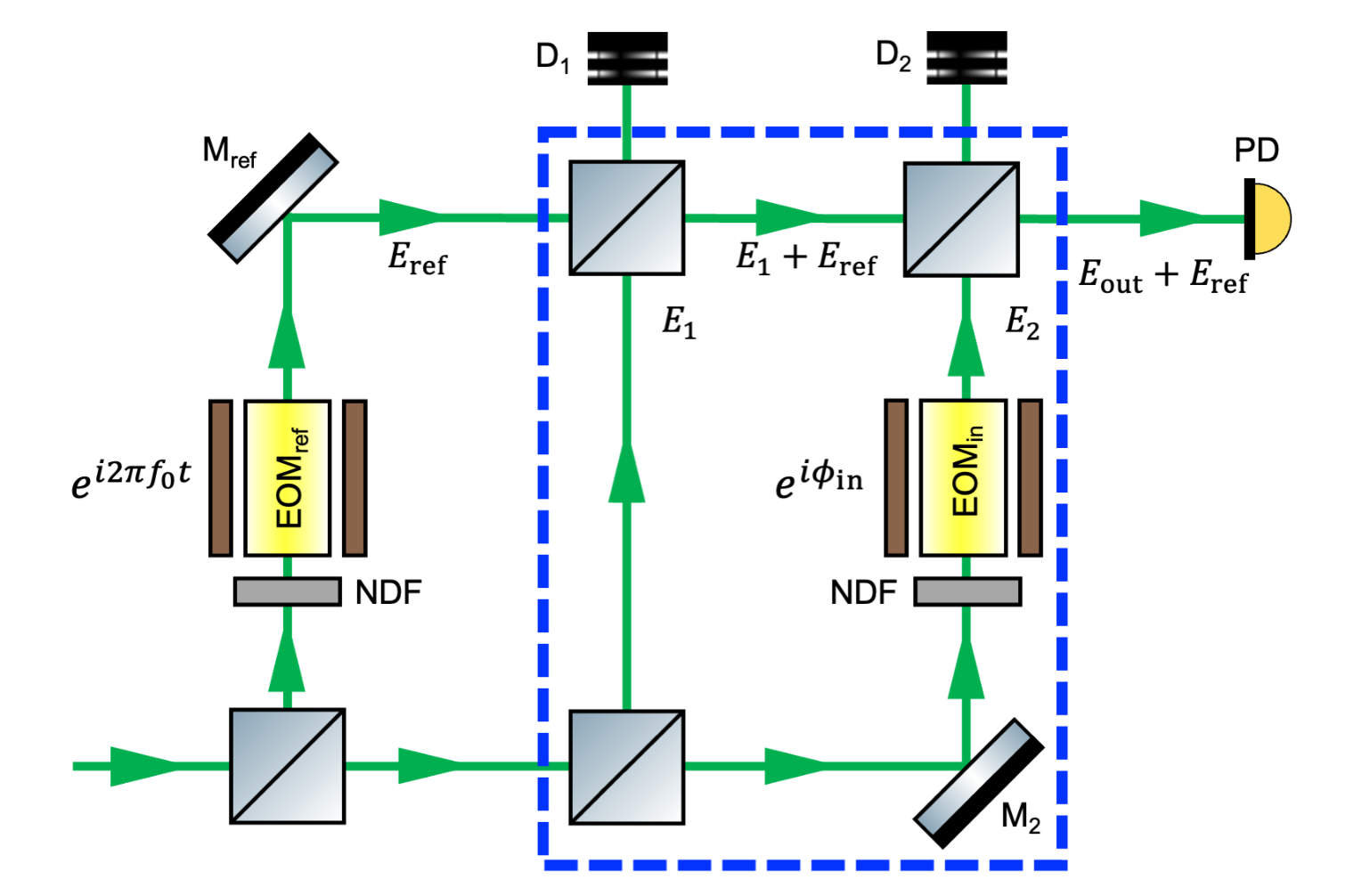}
\caption{Schematic of the experimental setup. The Mach--Zehnder interferometer (MZI) used for phase magnification is enclosed by the dashed blue rectangle. Its two arms provide the fields $E_1$ and $E_2$, which are nearly in phase opposition at the output, so that $E_{\textrm{out}}= E_1+E_2$ is the dark-port field. The leftmost branch, external to the MZI, provides the frequency-shifted time-holographic reference $E_{\textrm{ref}}$, which is combined with $E_{\textrm{out}}$ at the photodetector. The symbols $E_1$, $E_2$, and $E_{\textrm{ref}}$ indicate field paths; in the main text, they denote the corresponding field values at the photodetector. Labels: EOM, electro-optic modulator (Thorlabs EO-PM-NR-C4); PD, photodetector (Thorlabs DET100A2); NDF, variable neutral-density filter; D\textsubscript{1}, D\textsubscript{2}, beam dumps (D\textsubscript{1} terminates the reference--$E_1$ beat, available for a reference phase lock; D\textsubscript{2}, the complementary bright port); M\textsubscript{ref}, M\textsubscript{2}, mirrors (M\textsubscript{2}, piezo-actuated). The divided squares ($\boxslash$) represent 50/50 nonpolarizing beamsplitter cubes.}\label{fig:setup}
\end{figure}

We investigate phase magnification in a Mach--Zehnder interferometer (MZI), schematically shown in Fig.~\ref{fig:setup}. The light source is a 300~mW frequency-doubled single-mode laser at 532~nm (Spectra-Physics Excelsior 532-300-CDHR), having a spectral linewidth below $10~\textrm{MHz}$ ($<0.01~\textrm{pm}$; relative linewidth $<2\times10^{-8}$) and optical intensity noise below $0.2\%$ rms over $20~\textrm{Hz}$--$20~\textrm{MHz}$ (integrated RIN $<2\times10^{-3}$). Before entering the lower-left input beamsplitter, the laser beam passes through a half-wave plate (Thorlabs WPHSM05-532) and a Glan--laser polarizer (Thorlabs GL10-A), which together attenuate the beam and purify its polarization. These elements are not shown in Fig.~\ref{fig:setup}. All fields in the setup share the same linear polarization.

The piezo-actuated mirror M\textsubscript{2} is used to establish destructive interference between the two MZI arms at the dark port. The internal interferometer phase, $\phi_{\textrm{in}}$, is defined relative to this nominal dark-fringe condition and is controlled by electro-optic phase modulator EOM\textsubscript{in}. Besides applying arbitrary phase waveforms, EOM\textsubscript{in} provides a controllable bias phase for fine adjustment of the dark-fringe operating point. A variable neutral-density filter in arm 2 sets the small intensity imbalance between the interfering fields.

In our time-holographic implementation, the MZI output field $E_{\textrm{out}}$ is mixed at the photodetector with a coherent reference field $E_{\textrm{ref}}$ derived from the same laser. The reference is frequency shifted by $f_0$ using phase modulator EOM\textsubscript{ref}, driven by a sawtooth voltage leading to $2\pi$ modulation depth \cite{Cumming-1957,Johnson-2010}. Its intensity is adjusted with a variable neutral-density filter. As indicated in Fig.~\ref{fig:setup}, the holographic reference is injected through the upper-left nonpolarizing beamsplitter and shares the upper path of the interferometer before being recombined with the dark-port field at PD. The detector delivers a photocurrent
$i_{\textrm{holo}}(t)=\eta q\,I_{\textrm{holo}}(t)$, where $\eta$ is the detector
quantum efficiency, $q$ the electron charge, and
$I_{\textrm{holo}}(t)=\left|E_{\textrm{out}}(t)+E_{\textrm{ref}}(t)\right|^2/\kappa^2$
the incident photon flux; the latter constitutes the time hologram. Recalling that
$E_{\textrm{out}}(t)=\kappa A_{\textrm{out}}\textrm{e}^{\textrm{i}(\omega t+\phi_{\textrm{out}})}$ and
$E_{\textrm{ref}}(t)=\kappa A_{\textrm{ref}}\,\textrm{e}^{\textrm{i}(\omega_{\textrm{ref}}t+\phi_{\textrm{ref}})}$,
where $\omega_{\textrm{ref}}\equiv\omega+2\pi f_0$ and $f_0$ is the frequency shift, and introducing the slowly varying complex amplitude
\begin{subequations}
\begin{gather}\label{E-signal}
\mathcal{I}_{\textrm{signal}}=A_{\textrm{ref}}A_{\textrm{out}}\textrm{e}^{\textrm{i}(\phi_{\textrm{out}}-\phi_{\textrm{ref}})}\equiv \vert\mathcal{I}_{\textrm{signal}}\vert\,\textrm{e}^{\textrm{i}\phi_{\textrm{signal}}},
\end{gather}
the time hologram reads as
$I_{\textrm{holo}}(t)=I_{\textrm{ref}}+I_{\textrm{out}}+I_{\textrm{signal}}(t)$, where
\begin{gather}
        I_{\textrm{signal}}(t)\equiv \mathcal{I}_{\textrm{signal}}\textrm{e}^{-\textrm{i}2\pi f_0t}+ \mathcal{I}^\ast_{\textrm{signal}}\textrm{e}^{\textrm{i}2\pi f_0t}.\label{I-signal}
\end{gather}
\end{subequations}
Here, $I_{\textrm{ref}}\equiv A_{\textrm{ref}}^2$ is the reference intensity, $I_{\textrm{out}}$ is the conventional MZI output intensity [see Eq.~(\ref{I-Hecht})], and $I_{\textrm{signal}}(t)$ is the temporal-interference, or beating, term, where the relevant information is encoded.

\subsection{Complex-field readout}
The reference and output intensity terms, $I_{\textrm{ref}}$ and $I_{\textrm{out}}$, are relatively slow terms whose spectral support essentially lies around zero frequency. By contrast, the beating term $I_{\textrm{signal}}(t)$~(\ref{I-signal}) oscillates at the reference-shift frequency $f_0$ and therefore appears around $\pm f_0$ in the spectrum. Thus, by choosing $f_0$ appropriately, $I_{\textrm{signal}}(t)$ can be separated from the lower-frequency contributions and efficiently extracted from $I_{\textrm{holo}}(t)$ by Fourier demodulation \cite{Takeda-1982}. This procedure yields the complex quantity $\mathcal{I}_{\textrm{signal}}$~(\ref{E-signal}), essentially the complex dark-port field referred to the holographic reference.

The sum $I_{\textrm{ref}}+I_{\textrm{out}}\simeq I_{\textrm{ref}}$ forms a slow pedestal that carries no phase information yet is much larger than the beating term. In the experiment reported here the full hologram, pedestal included, is digitized and recorded. A simple improvement, not implemented here, is to remove the pedestal with an analog high-pass filter before recording, so that the digitizer captures essentially the beat photocurrent $i_{\textrm{signal}}(t)=\eta q\,I_{\textrm{signal}}(t)$ at $f_0$; this preserves the dynamic range of the analog-to-digital conversion without affecting the beating term or the retrieved phase, as discussed in Supplement 1, Sec.~S3.4.

In either case, after demodulation, the two quadratures of $\mathcal{I}_{\textrm{signal}}$ give the tangent of the output phase directly,
\begin{equation}
T_{\textrm{signal}}\equiv
\frac{\operatorname{Im}\mathcal{I}_{\textrm{signal}}}
{\operatorname{Re}\mathcal{I}_{\textrm{signal}}}
=\tan\phi_{\textrm{signal}}=\tan(\phi_{\textrm{out}}-\phi_{\textrm{ref}}).
\label{T-holo}
\end{equation}
Being a ratio of the two quadratures, $T_{\textrm{signal}}$ is independent of the overall detection gain $\eta q$ and of any common-mode intensity fluctuation. The reference $\phi_{\textrm{ref}}$ appears as a phase offset in $T_{\textrm{signal}}$. Here, this offset is determined from a fit to the expected linear relation, Eq.~(\ref{lin-tan}), written as
\begin{equation}\label{fit-0}
\tan(\phi_{\textrm{signal}}+\phi_{\textrm{ref}})=M\phi_{\textrm{in}},
\end{equation}
where $\phi_{\textrm{signal}}=\arctan T_{\textrm{signal}}$. This strategy is possible because, in the present proof-of-principle experiment, $\phi_{\textrm{in}}$ is a controlled variable and the aim of the work is to demonstrate and calibrate the phase-magnification effect.

When the same principle is used instead for metrology, i.e.\ to infer an unknown $\phi_{\textrm{in}}$, the reference offset $\phi_{\textrm{ref}}$ must be independently determined and deducted. A natural way to do this within our setup is to actively stabilize $\phi_{\textrm{ref}}$ by means of a phase-lock loop acting either on the position of mirror M\textsubscript{ref}, using a piezoelectric transducer, or on EOM\textsubscript{ref}. The error signal can be obtained from the presently unused output beam terminating in beam dump D\textsubscript{1} in Fig.~\ref{fig:setup}. After this phase correction, $T_{\textrm{signal}}$ essentially gives the desired estimator $T_{\textrm{out}}$.

\subsection{The imbalance parameter}
The imbalance $\mu$ is fixed from power measurements by any of three related routes. From the arm powers, $A_1/A_2=\sqrt{P_1/P_2}$ in Eq.~(\ref{def:mu}) gives $\mu=1-\sqrt{P_1/P_2}$. From the dark port, near the dark fringe Eqs.~(\ref{I-Hecht}) and (\ref{def:mu}) give, to leading order,
\begin{equation}\label{P_out_gen}
\frac{P_{\textrm{out}}}{P_2}\simeq\frac{P_{\textrm{out}}}{P_1}\simeq \mu^2+\phi_{\textrm{in}}^2,
\end{equation}
so the minimum output power yields
\begin{equation}\label{mu2-exp}
\mu^2=4\frac{P_{\textrm{out}}^{\textrm{min}}}{P_{\textrm{in}}},\qquad P_{\textrm{in}}\equiv 2(P_1+P_2),
\end{equation}
with $P_{\textrm{in}}$ the effective power injected into the MZI---exactly the total output $P_{\textrm{out}}+P_{\textrm{bright}}$, where the bright-port power $P_{\textrm{bright}}$ is given by Eq.~(\ref{I-Hecht}) with $+$ in place of $-$ (this route fixes $|\mu|$, not its sign). Finally, the dark-port route~(\ref{mu2-exp}) admits a self-normalizing realization through the complementary bright-port output, currently dumped at $D_2$ (Fig.~\ref{fig:setup}): the two ports satisfy $P_{\textrm{out}}/P_{\textrm{bright}}\simeq(\mu^2+\phi_{\textrm{in}}^2)/4$ to leading order, with $P_{\textrm{bright}}\simeq P_{\textrm{in}}$, so the minimum gives $\mu^2=4P_{\textrm{out}}^{\textrm{min}}/P_{\textrm{bright}}$---literally Eq.~(\ref{mu2-exp}) with $P_{\textrm{bright}}$ in place of $P_{\textrm{in}}$.

These routes differ in accuracy: the dark-port determination---in either its direct~(\ref{mu2-exp}) or ratiometric form---is the most accurate, and its precision, which caps the largest magnification that can be reliably fixed, is ultimately limited by the stability of the dark-fringe operating point rather than by the photon count (Supplement 1, Sec.~S2).

\subsection{Shot-noise limit to phase magnification holographic readout}

What ultimately bounds the resolution is not the magnification but the photon statistics of the readout: once technical disturbances are controlled, the floor is the shot noise of the detected light---the standard quantum limit of phase estimation with coherent states~\cite{Caves1981}, the limit that squeezed-light interferometers push against~\cite{Jia-2024}. We summarize how it fixes the minimum resolvable phase change $\sigma_{\phi_{\textrm{in}}}$---the rms uncertainty of ${\phi_{\textrm{in}}}$---deriving it fully in Supplement 1, Sec.~S3.

From Eq.~(\ref{lin-tan}), $\sigma_{\phi_{\textrm{in}}}=\vert\mu\vert\,\sigma_{T_{\textrm{out}}}$ (Supplement 1, Sec.~S1). The quantity actually measured, however, is not $T_{\textrm{out}}$ but $T_{\textrm{signal}}$ [Eqs.~(\ref{T-holo}),~(\ref{E-signal})], which carries the dark-port phase only relative to the reference phase $\phi_{\textrm{ref}}$. The latter is fixed with negligible uncertainty---here from the fit of Eq.~(\ref{fit-0}) to the controlled $\phi_{\textrm{in}}$, and by an active reference lock for metrology---so that $T_{\textrm{signal}}\simeq T_{\textrm{out}}$ and all genuine measurement uncertainty resides in the dark-port quadratures, $\textrm{Re}\,\mathcal{I}_{\textrm{signal}}$ and $\textrm{Im}\,\mathcal{I}_{\textrm{signal}}$, recovered by Fourier demodulation of the time hologram around $f_0$~\cite{Takeda-1982}.

In the heterodyne regime ($I_{\textrm{ref}}\gg I_{\textrm{out}}$) the reference acts as a strong local oscillator, setting a shot-noise level that is isotropic and independent in the two quadratures~\cite{Haus2000,BachorRalph2019,SalehTeich2019}. Decisively, $T_{\textrm{signal}}$ is a \emph{ratio} of these quadratures, so any common-mode fluctuation---above all the laser relative-intensity noise shared by the reference and dark-port fields---divides out exactly, leaving shot noise as the fundamental limit~\cite{Caves1981}: $\sigma^2_{\phi_{\textrm{signal}}}\simeq{1}/({2N_{\textrm{out}}})$, with $N_{\textrm{out}}\equiv I_{\textrm{out}}\tau$ the mean dark-port photon count over the effective integration time $\tau\equiv1/\Delta f$ fixed by the equivalent noise bandwidth $\Delta f$ of the demodulation window~\cite{Harris-1978,Kingston-book}. The reference power thus cancels---it jointly sets the gain and the shot level, the canonical heterodyne result~\cite{Haus2000}---and the resolution depends solely on the dark-port flux (Supplement 1, Sec.~S3.1). Propagating this uncertainty through Eq.~(\ref{T-holo}) gives
\begin{equation}\label{sigma_Tout}
\sigma_{T_{\textrm{out}}}=\frac{1+T_{\textrm{out}}^2}{\sqrt{2N_{\textrm{out}}}},
\end{equation}
after using $\sigma_{\phi_{\textrm{ref}}}\ll\sigma_{\phi_{\textrm{signal}}}$ as already announced. Now we can use $\sigma_{\phi_{\textrm{in}}}=\vert\mu\vert\,\sigma_{T_{\textrm{out}}}$. Recalling Eq.~(\ref{P_out_gen}) with $\phi_{\textrm{in}}=\mu T_{\textrm{out}}$, and $P_2\simeq P_{\textrm{in}}/4$, yields $\mu^2(1+T_{\textrm{out}}^2)\simeq4P_{\textrm{out}}/P_{\textrm{in}}=4N_{\textrm{out}}/N_{\textrm{in}}$, with $N_{\textrm{in}}\equiv I_{\textrm{in}}\tau$ the photons injected into the MZI during $\tau$ ($I_{\textrm{in}}\equiv P_{\textrm{in}}/\hslash\omega$ being the injected photon flux). This result along with Eq.~(\ref{sigma_Tout}) eliminates $\mu$ explicitly and yields
\begin{equation}\label{sigma_phiin_SQL}
\sigma_{\phi_{\textrm{in}}}=\sqrt{\frac{2\left(1+T_{\textrm{out}}^2\right)}{N_{\textrm{in}}}}.
\end{equation}
The factor of two is the so-called 3~dB heterodyne penalty~\cite{YuenChan1983,Haus2000}, the unavoidable cost of measuring both quadratures at once---that is, of the complex-field readout itself (Supplement 1, Sec.~S3.1). The magnification survives only implicitly, via $T_{\textrm{out}}=M\phi_{\textrm{in}}$, and the two regimes pull oppositely. At the true dark fringe ($\phi_{\textrm{in}}^2\ll\mu^2$, i.e.\ $\vert T_{\textrm{out}}\vert\ll1$) the floor tends to $\sqrt{2/N_{\textrm{in}}}$, independent of the chosen $\mu$, so magnifying more costs nothing. Farther out ($\vert T_{\textrm{out}}\vert\gtrsim1$) it grows as $\vert T_{\textrm{out}}\vert\sqrt{2/N_{\textrm{in}}}$, so for a given excursion the magnification should be capped to keep $T_{\textrm{out}}=\mathcal{O}(1)$ if shot noise is the limiting factor.

Numbers make the point. For typical values of our setup, $P_{\textrm{in}}\sim10~\textrm{mW}$, $\lambda=532~\textrm{nm}$, and $\Delta f=200~\textrm{Hz}$ ($\tau\approx5~\textrm{ms}$), the dark-fringe floor is $\sigma_{\phi_{\textrm{in}}}\approx0.12~\mu\textrm{rad}$---equivalent to an optical-path variation of about $10~\textrm{fm}$. Away from the dark fringe, it increases only by the factor $\sqrt{1+T_{\textrm{out}}^2}$ and remains below the microradian level over the range of output tangents used here. This is far below the resolution demonstrated in Sec.~\ref{sec:exp-results}. The present proof-of-principle is therefore limited by uncontrolled environmental phase drift, rather than by shot noise; residual laser phase noise is likewise subdominant in this common-laser configuration. Further lowering the floor is then a matter of photon budget: increasing the intra-interferometer power---the 300-mW source leaves more than an order of magnitude in reserve---or, where the signal bandwidth permits, reducing $\Delta f$, at the expense of temporal resolution. Conversely, the $P_{\textrm{in}}^{-1/2}$ scaling leaves ample headroom at low power: even at $P_{\textrm{in}}\sim10~\mu\textrm{W}$---of interest for photosensitive samples---the dark-fringe floor is only $\approx4~\mu\textrm{rad}$ ($\approx0.3~\textrm{pm}$), still more than an order of magnitude below the resolution demonstrated here.

\section{Experimental results}\label{sec:exp-results}

\subsection{Environmental conditions}

The setup was not actively stabilized, and no special effort was made to isolate it from environmental disturbances beyond the use of a pneumatic optical table. Only a mild control of the bias voltage applied to EOM\textsubscript{in} was performed between measurements, in order to compensate for slow drifts of the dark fringe. Under these minimal isolation conditions, unwanted phase fluctuations, including slow drifts induced by air currents and airborne vibrations, can affect individual records and lead to significant noise. This noise was worsened by intermittent construction-related vibrations on campus during the measurement campaign. Since the experiments typically operate with phase magnifications of order $10^3$, optical-path variations of only a few tens of picometers over the acquisition time can noticeably compromise the retrieved signal. Measurement records dominated by uncontrolled phase variations---typically in the 10~Hz range, close to the minimum resolvable frequency; see below---were therefore discarded. This limitation is not intrinsic to the method, but reflects the environmental isolation conditions of the present proof-of-principle implementation. Nevertheless, for the remaining records, standard noise-subtraction methods \cite{Boll-1979} are effective, as shown below, and this noise does not impose a major limitation on the approach. A detailed description of the full procedure is provided in Supplement 1, Sec.~S4.

\begin{figure}[!ht]
\centering\includegraphics[width=0.7\linewidth]{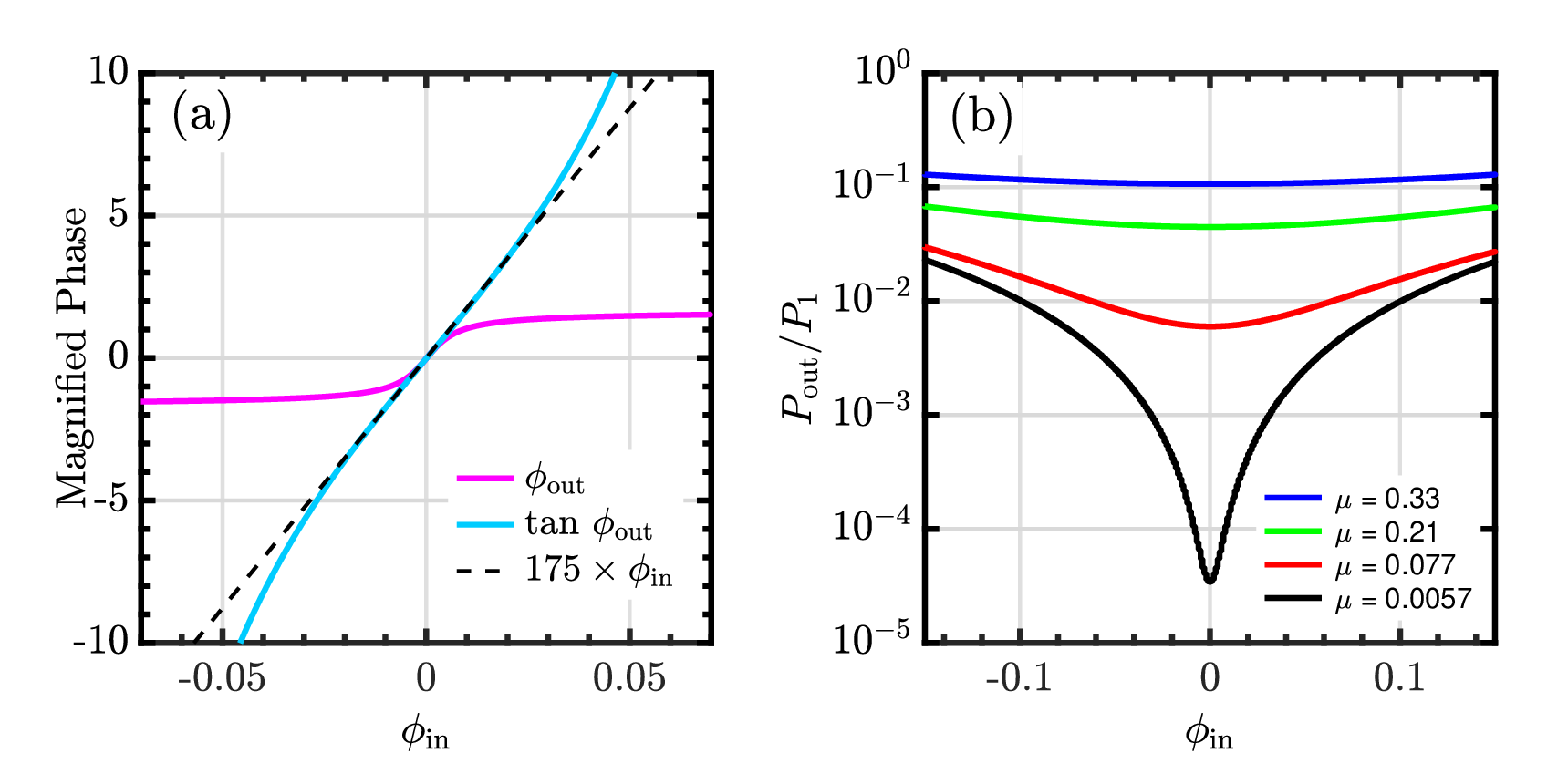}
\caption{Experimentally determined phase magnification and output power as functions of the internal phase $\phi_{\textrm{in}}$. (a) Output phase $\phi_{\textrm{out}}$ and its tangent $\tan\phi_{\textrm{out}}$ for an imbalance $\mu\approx0.0057$ ($M\approx175$), together with the ideal magnified signal $175\,\phi_{\textrm{in}}$ (dashed line). (b) Normalized output power $P_{\textrm{out}}/P_1$, on a logarithmic scale, for four values of the imbalance parameter $\mu$ (legend): smaller $\vert\mu\vert$ produces deeper output minima and a steeper phase response. The lowest, black curve in (b) corresponds to the case depicted in (a).}\label{fig:sigmoid-exp}
\end{figure}

\subsection{Results}\label{sec:exp}

To demonstrate the effect, we drive the electro-optic modulator EOM\textsubscript{in} with different temporal waveforms repeated every 120~ms, which defines the acquisition window of each temporal hologram. This repetition period sets the minimum resolvable frequency to 8.33~Hz. The frequency offset between the holographic reference and the central frequency of the MZI output field is set to $f_0=4\,\textrm{kHz}$, obtained by driving EOM\textsubscript{ref} with a sawtooth voltage of modulation depth $V_{2\pi}$ and period 0.25~ms. We use several input phase waveforms $\phi_{\mathrm{in}}(t)$, with different shapes and central frequencies around 200~Hz, well below $f_0$.

Figure~\ref{fig:sigmoid-exp} demonstrates the basic magnification effect in (a) and the relation between the darkness of the output and the imbalance in (b). Here, EOM\textsubscript{in} is driven with a triangular signal, producing an effectively step-wise linear variation of $\phi_{\textrm{in}}(t)$. The experiment is repeated for different imbalance parameters $\mu$, as indicated in the legend. Measurements of the output power $P_{\textrm{out}}$ and of the arm powers $P_{1,2}$---needed to determine $\mu$ through Eq.~(\ref{mu2-exp})---are performed by blocking the holographic reference and, when appropriate, one of the internal arms. The slope around the origin in Fig.~\ref{fig:sigmoid-exp}(a) increases as the imbalance $\mu$ decreases, in correspondence with the deeper output minima shown in Fig.~\ref{fig:sigmoid-exp}(b). This is the signature of phase magnification, which in this case reaches $M\approx175$ for the steepest, black continuous curves; Supplement 1, Sec.~S5 characterizes this imbalance--magnification relationship for the full set of imbalances, showing the output phase alongside its tangent.

\begin{figure}[!ht]
\centering
\includegraphics[width=0.6\linewidth]{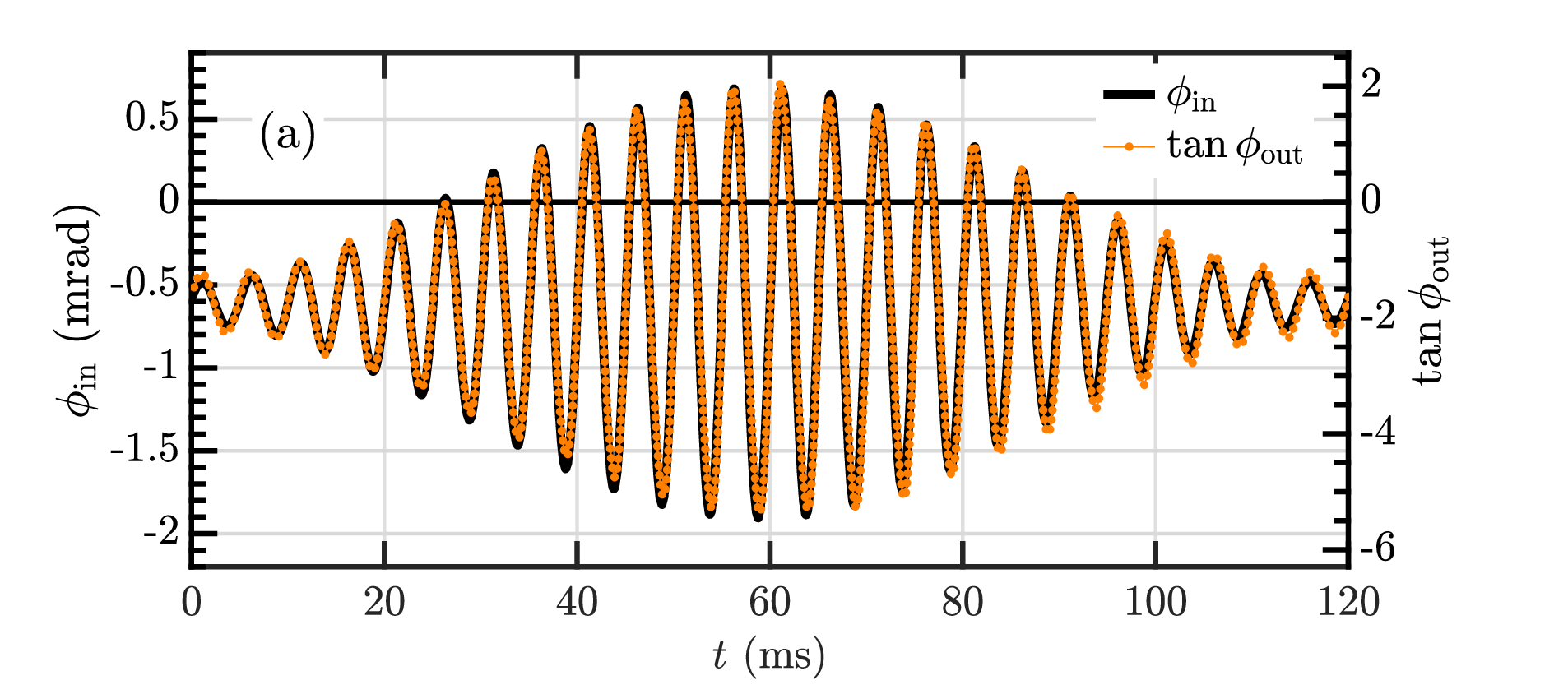}
\includegraphics[width=0.6\linewidth]{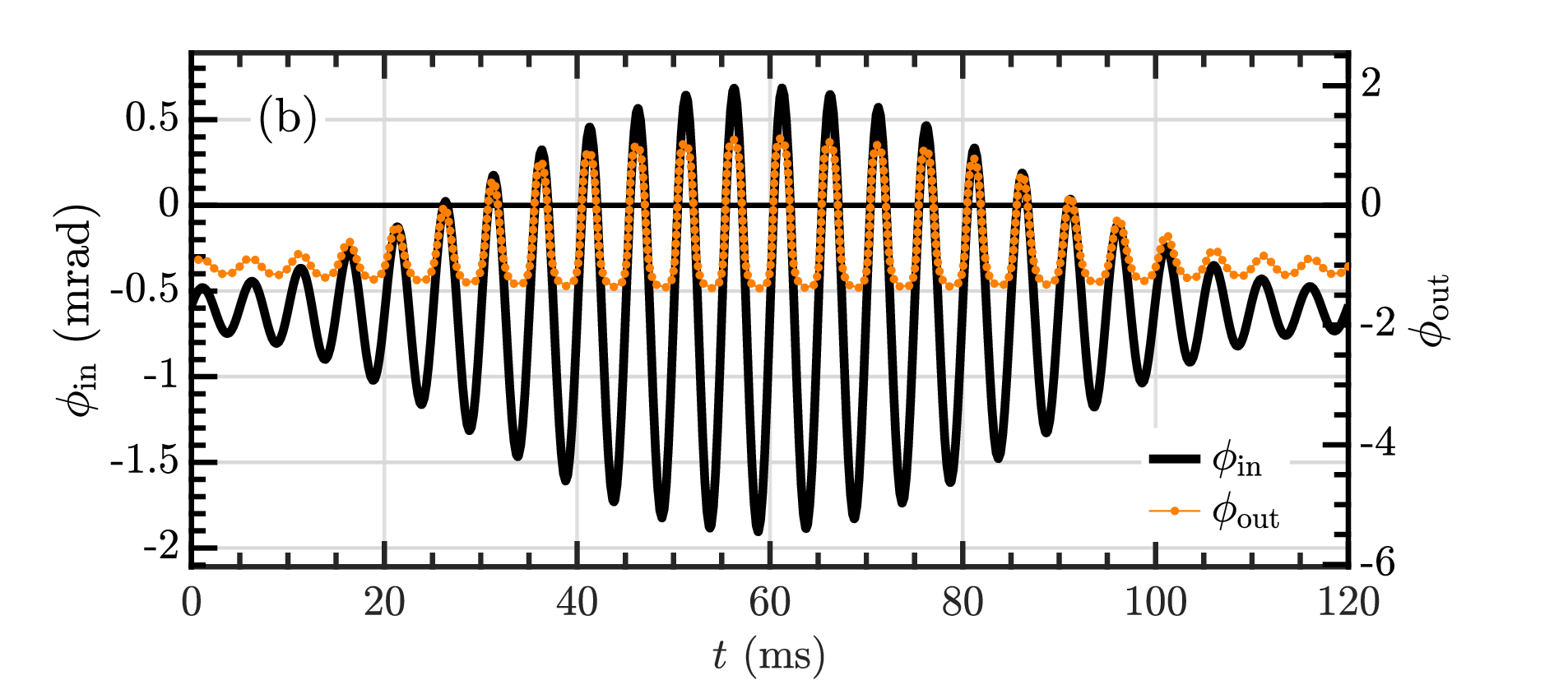}
\caption{Experimental magnification of a Gaussian-enveloped sinusoidal temporal phase signal with a maximum modulation depth of $2.5~\textrm{mrad}$ and a central frequency of $200~\textrm{Hz}$. In both panels, the solid black line is the input phase $\phi_{\textrm{in}}$ as a function of time $t$, referred to the left axis in mrad, and the orange dots are the retrieved output, referred to the right axis: the tangent $\tan\phi_{\textrm{out}}$ in (a) and the output phase $\phi_{\textrm{out}}$ in (b). The disparity between the left- and right-axis ranges expresses the achieved magnification, $M\approx2{,}900$. Whereas the tangent in (a) reproduces the input over the full excursion, $\phi_{\textrm{out}}$ in (b) saturates against it.}\label{fig:singauss-exp}
\end{figure}

\begin{figure}[!ht]
\centering
\includegraphics[width=0.6\linewidth]{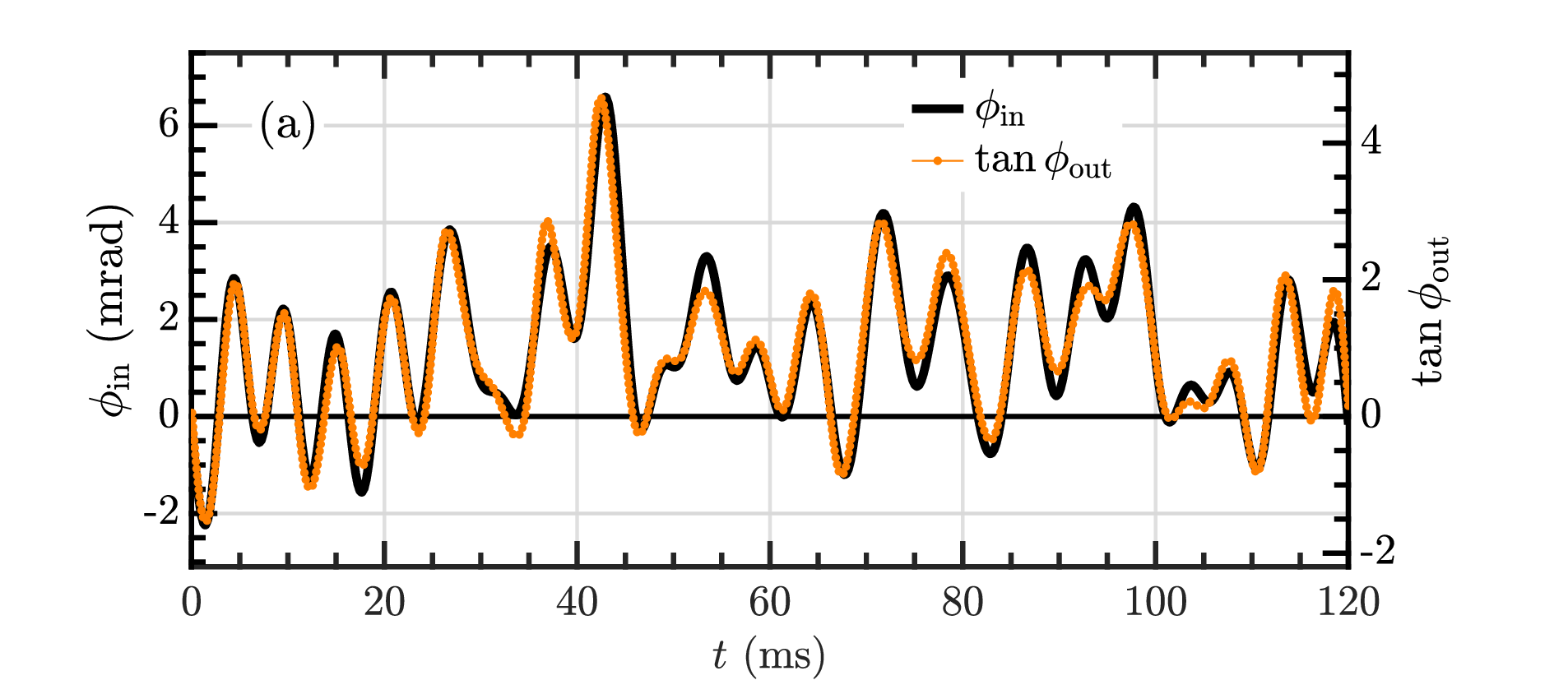}
\includegraphics[width=0.6\linewidth]{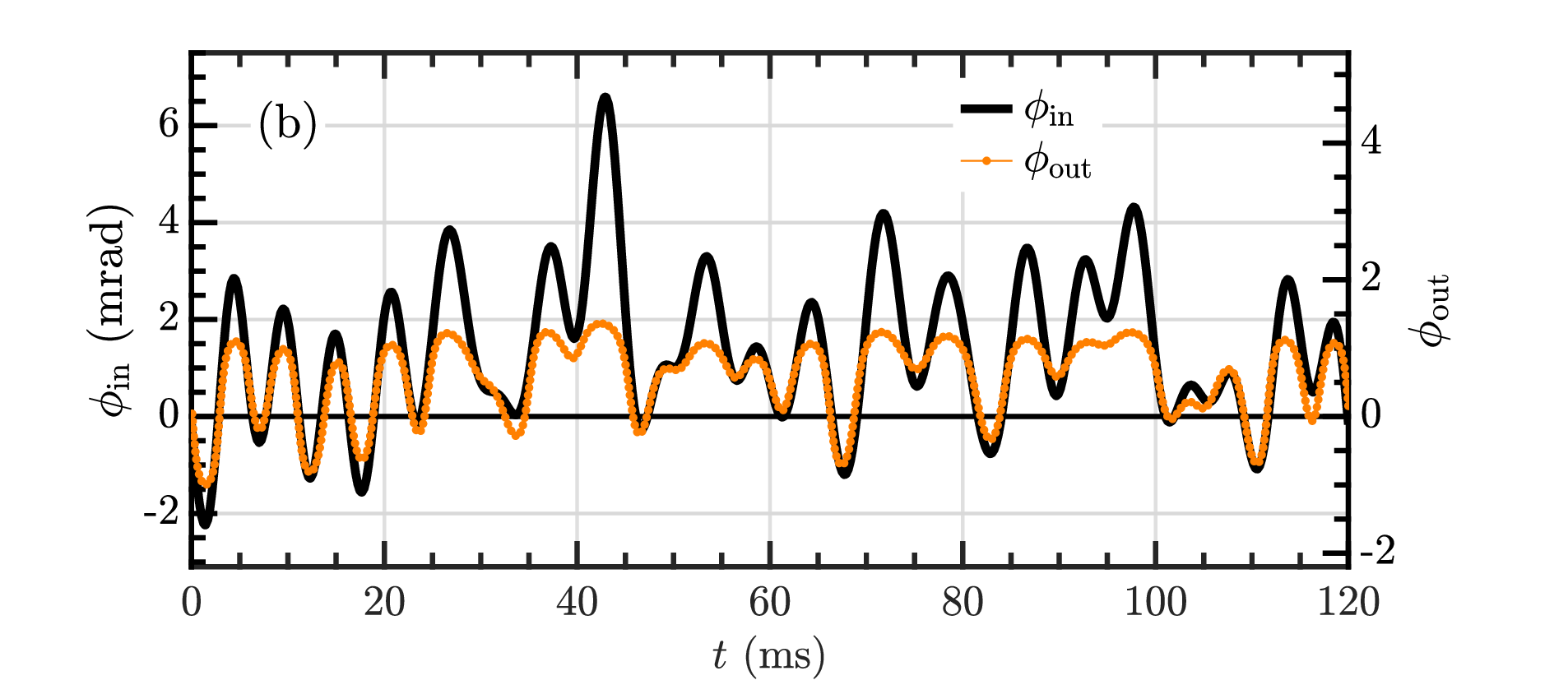}
\caption{Experimental magnification, with $M\approx710$, of a pseudo-random phase signal generated with a finite bandwidth of $200~\textrm{Hz}$, shown as a function of time $t$; same conventions as in Fig.~\ref{fig:singauss-exp}. In (a) the tangent estimator $\tan\phi_{\textrm{out}}$ compares favorably with the signal, while in (b) the phase estimator $\phi_{\textrm{out}}$ saturates against the larger excursions, as in Fig.~\ref{fig:singauss-exp}(b).}\label{fig:speckle2-exp}
\end{figure}

A main goal of this work is to demonstrate the magnification of temporal phase signals displaying sub-mrad features---corresponding to optical paths in the $10~\textrm{pm}$ range---embedded in larger excursions in the mrad scale, sub-nm range. This is exemplified in Figs.~\ref{fig:singauss-exp} and \ref{fig:speckle2-exp}, which correspond to two qualitatively different scenarios: a Gaussian-enveloped sinusoidal signal and a pseudo-random signal, respectively. In both cases, the continuous black line represents the MZI internal phase $\phi_{\textrm{in}}(t)$, referred to the left axis in mrad, while the orange dots give the experimentally retrieved $\tan\phi_{\textrm{out}}(t)$ in (a) or $\phi_{\textrm{out}}(t)$ in (b), referred to the right axis. The very different spans of the two vertical axes directly reveal the large phase magnification $M$, with values around $2{,}900$ in Fig.~\ref{fig:singauss-exp} and $710$ in Fig.~\ref{fig:speckle2-exp}. The output phase was determined by fitting Eq.~(\ref{fit-0}), with $\phi_{\textrm{in}}(t)=\phi_{\textrm{in},0}+\phi_{\textrm{eom}}(t)$, where $\phi_{\textrm{eom}}(t)$ is the phase change imparted by EOM\textsubscript{in} and $\phi_{\textrm{in},0}$ is a residual offset from the dark fringe, which cannot be controlled at this level. We find $\phi_{\textrm{in},0}\approx-0.61~\textrm{mrad}$ and $+1.30~\textrm{mrad}$ in Figs.~\ref{fig:singauss-exp} and \ref{fig:speckle2-exp}, respectively, corresponding to residual optical-path differences between the MZI arms of $\approx-52$~pm and $+110~\textrm{pm}$.

The figures clearly show that using $\tan\phi_{\textrm{out}}$ instead of $\phi_{\textrm{out}}$ considerably improves the method. Indeed, the results are consistent with our theoretical analysis: for $M=2{,}900$ [Fig.~\ref{fig:singauss-exp}], the linear range of $\phi_{\textrm{out}}$ as an estimator of $\phi_{\textrm{in}}$ is about $\vert\phi_{\textrm{in}}\vert\lesssim0.14~\textrm{mrad}$ [Eq.~(\ref{lin-Ariel})], far below the dynamic range in this case, as seen in Fig.~\ref{fig:singauss-exp}(b); note the horizontal line marking the zero. This explains the evident saturation of the estimator against relatively large internal phase excursions; see Fig.~\ref{fig:sigmoid-theory}. By contrast, using $\tan\phi_{\textrm{out}}$ as an estimator extends the linear range to $\vert\phi_{\textrm{in}}\vert\lesssim5.6~\textrm{mrad}$ [Eq.~(\ref{lin-tan})], easily accommodating the phase excursion despite the imperfectly determined zero, as seen in Fig.~\ref{fig:singauss-exp}(a).

Inspection of Fig.~\ref{fig:singauss-exp}(a) further shows that the method resolves $\sim10~\textrm{pm}$ optical-path variations ($\sim0.1~\textrm{mrad}$) embedded into $\sim0.2~\textrm{nm}$ excursions ($\sim2.5~\textrm{mrad}$). In fact, the rms deviation of the experimental points from the input values in Fig.~\ref{fig:singauss-exp}(a) evaluates to $\sigma_\phi=0.06~\textrm{mrad}$, corresponding to $5~\textrm{pm}$.

Experiments with a pseudo-random signal [Fig. \ref{fig:speckle2-exp}] resulted in more noisy outputs due to the daily variation of environmental conditions and the intrinsic nature of the signal. Our experiments show an average error of $0.34~\textrm{mrad}$, equivalent to $29~\textrm{pm}$. Note that, in the presence of strong environmental noise, broadband signals are inherently noisier than narrow-band ones, as the noise fully overlaps the signal across its bandwidth, making noise reduction by spectral subtraction less efficient.

\section{Discussion and outlook}

Conventional interferometry relies on power readout, for which perfectly balanced interferometers are traditionally regarded as optimal because they maximize fringe contrast. The situation changes markedly when the phase of the field leaving the interferometer is read out instead. In that case, a marginal imbalance between the two arms---quantified by the parameter $\mu$ ($\vert\mu\vert\ll1$)---produces a dark-port phase $\phi_{\textrm{out}}$ that is extremely sensitive to small internal phase variations $\phi_{\textrm{in}}$ around the point of minimum output power, i.e., the dark fringe. Indeed we have shown that $T_{\textrm{out}}\equiv\tan\phi_{\textrm{out}}\simeq M\phi_{\textrm{in}}$, with magnification $M\equiv\mu^{-1}$ ($\vert{M}\vert\gg1$), over a linear dynamic range that scales as $\vert\mu\vert^{1/2}$; see Eq.~(\ref{lin-tan}).

Two key elements underpin this approach. First, $T_{\textrm{out}}$ is the natural quantity to read, because near the dark fringe the dark-port response is linear in the tangent [Eq.~(\ref{lin-tan})]. The output phase $\phi_{\textrm{out}}=\arctan(M\phi_{\textrm{in}})$---the estimator of Refs.~\cite{Ariel-2018,Ariel-2020}---saturates instead: its usable range is smaller by a factor $\sim\sqrt{\vert\mu\vert}\ll1$, and the saturation is that of the arctangent, not of the interferometer [Fig.~\ref{fig:sigmoid-theory}]. Using $T_{\textrm{out}}$ thus gives access to fine substructure within phase excursions that, although still very small in absolute terms, would already exceed the linear range of the phase estimator. Second, the dark-port complex amplitude is retrieved by time-domain holography, through its interference with a monochromatic, frequency-shifted reference. Complex-field retrieval is what distinguishes this readout from the phase-shift-amplified schemes of Refs.~\cite{Ariel-2018,Ariel-2020}, which access the magnified phase through a single projection---the detected power~\cite{Ariel-2018} or the cosine of the beat phase after synchronous detection~\cite{Ariel-2020}---and whose immunity to laser relative-intensity noise (RIN) grows with the magnification itself~\cite{Ariel-2020}. With both quadratures available, the estimator $T_{\textrm{out}}$ is their ratio: sign-resolved, free of amplitude calibration, and immune to RIN at any magnification, so that noise immunity is decoupled from gain and $M$ is freed to serve dynamic range alone. These developments build on earlier exploitation of the same steep dark-port response along two independent lines: the phase-shift-amplified interferometry pioneered by Sternklar and co-workers, which first turned imperfect destructive interference into a practical sensitivity resource~\cite{Ariel-2015,Ariel-2018,Ariel-2020}, and our own use of its near-discontinuous limit to detect phase-sign changes~\cite{OTD}.

In a proof-of-principle tabletop Mach--Zehnder interferometer (MZI) experiment [Fig.~\ref{fig:setup}], performed without optimization or specific environmental isolation, we have demonstrated phase magnification factors of order $10^3$ and resolved optical-path variations down to 5~pm, better than $\lambda/100{,}000$, as shown in Fig.~\ref{fig:singauss-exp}(a). Figures~\ref{fig:singauss-exp}(b) and \ref{fig:speckle2-exp}(b) show the corresponding saturation of $\phi_{\textrm{out}}$, with the fine substructure nonetheless accessible through the tangent readout. The time-holographic readout was implemented by electro-optic serrodyne frequency shifting of the reference beam, placing the heterodyne offset in the audio band ($f_0=4~\textrm{kHz}$), with the recorded hologram Fourier-demodulated over millisecond windows. The shot-noise floor sits a factor of 90--500 below this environment-limited resolution (Supplement 1, Sec.~S3.3).

The principle demonstrated here opens several directions for further development. The most immediate one is conceptually straightforward, although experimentally demanding: improved environmental isolation and active stabilization, which should plausibly push the quantitative resolution into the sub-pm range. On the readout side, replacing the electro-optic modulators with acousto-optic ones would carry the heterodyne offset $f_0$ into the MHz range---above the laser's relaxation-oscillation band---lifting the demodulation band clear of the $1/f$ and audio-band technical noise without affecting the shot-noise floor (Supplement 1, Sec.~S3.4). The laser source itself offers two further, independent levers, which follow from the floor and laser-phase-noise scalings of Supplement 1, Secs.~S3.3 and S3.4: a watt-class single-frequency source at 532~nm lowers the shot-noise floor by an order of magnitude---into the 10-nrad, femtometer class---at the modest cost of millimeter-level path matching, while a kHz-linewidth source relaxes path matching from centimeters to the meter scale, opening fiber-delivered and dissimilar-arm geometries. A class-A source, free of relaxation oscillations~\cite{Baili-2007}, additionally leaves the choice of $f_0$ essentially unconstrained. Beyond temporal phase sensing, the approach extends naturally to quantitative phase imaging of transparent objects that are either extremely optically thin or exhibit very fine transverse structure; in reflection geometries, it could likewise provide a complementary tool for ultrafine surface profilometry. Work is currently in progress on quantitative phase imaging and on exploring alternative interferometers for temporal phase-signal magnification.

\section*{Funding}
This work is part of Projects \seqsplit{PID2020-120056GB-C21}, \seqsplit{PID2020-120056GB-C22}, \seqsplit{PID2023-153363NB-C21}, and \seqsplit{PID2023-153363NB-C22}, funded by \seqsplit{MICIU/AEI/10.13039/501100011033}.

\section*{Acknowledgment}
We are grateful to Jos\'{e} \'{A}ngel Picazo-Bueno, Mart\'{i}n Sanz, Eugenio Rold\'{a}n, and Carlos Navarrete-Benlloch for valuable discussions and advice.

\section*{Disclosures}
The authors declare no conflicts of interest.

\section*{Data Availability Statement}
Data underlying the results presented in this paper are not publicly available at this time but may be obtained from the authors upon reasonable request.

\section*{Supplemental Document}
See Supplement 1 for supporting content: Sec.~S1, the propagation-of-uncertainties framework; Sec.~S2, the determination of the imbalance parameter; Sec.~S3, the shot-noise limit and noise budget of the holographic readout; Sec.~S4, the environmental-noise reduction procedure; and Sec.~S5, the experimental characterization of the imbalance--magnification relationship.


\begin{thebibliography}{99}
\bibitem{Hariharan-book} P. Hariharan, \emph{Optical Interferometry}, 3rd ed. (Academic Press, 2003).

\bibitem{Michelson} A. A. Michelson and E. W. Morley, \enquote{On the relative motion of the Earth and the luminiferous ether,} American Journal of Science \textbf{34}, 333--345 (1887).

\bibitem{Zehnder} L. Zehnder, \enquote{Ein neuer Interferenzrefraktor,} {Z. Instr.} \textbf{11}, 275--285 (1891).

\bibitem{Mach} L. Mach, \enquote{\"{U}ber einen Interferenzrefraktor,}
{Sitzungsber. Kaiserl. Akad. Wiss., Math.-Naturwiss. Cl.} \textbf{103}, 610--632 (1894).

\bibitem{Hecht-book} E. Hecht, \emph{Optics}, 4th ed. (Addison-Wesley, 2002), p. 282.

\bibitem{F-P} A. Perot and C. Fabry, \enquote{On the application of interference phenomena to the solution of various problems of spectroscopy and metrology,} Astrophysical Journal \textbf{9}, 87--115 (1899).

\bibitem{GTI} F. Gires and P. Tournois, \enquote{Interf\'{e}rom\`{e}tre utilisable pour la compression d'impulsions lumineuses modul\'{e}es en fr\'{e}quence,} C. R. Acad. Sci. Paris \textbf{258}, 6112-6115 (1964).

\bibitem{Harimoto-2023} T. Harimoto, \enquote{Enhanced phase-shifting measurement with a double-passed Michelson interferometer,} Optical Review \textbf{30}, 331--335 (2023).

\bibitem{Ren-2008}
D. Ren, K. M. Lawton, and J. A. Miller,
\enquote{A double-pass interferometer for measurement of dimensional changes,}
Measurement Science and Technology \textbf{19}, 025303 (2008).

\bibitem{Schwarze-2023}
C. R. Schwarze, D. S. Simon, and A. V. Sergienko, \enquote{Enhanced-sensitivity interferometry with phase-sensitive unbiased multiports,} Physical Review A \textbf{107}, 052615 (2023).

\bibitem{GW} B. P. Abbott, R. Abbott, T. D. Abbott \emph{et al.} (LIGO Scientific Collaboration and Virgo Collaboration), \enquote{Observation of gravitational waves from a binary black hole merger,} Physical Review Letters \textbf{116}, 061102 (2016).

\bibitem{Hild-2009} S. Hild, H. Grote, J. Degallaix \emph{et al.},
\enquote{DC-readout of a signal-recycled gravitational wave detector,} Classical and Quantum Gravity \textbf{26}, 055012 (2009).

\bibitem{Fricke-2012} T. T. Fricke, N. D. Smith-Lefebvre, R. Abbott, \emph{et al.},
\enquote{DC readout experiment in Enhanced LIGO,} Classical and Quantum Gravity \textbf{29}, 065005 (2012).

\bibitem{Jia-2024} W. Jia, V. Xu, K. Kuns, \emph{et al.}, \enquote{Squeezing the quantum noise of a gravitational-wave detector below the standard quantum limit,}  Science \textbf{385}, 1318--1321 (2024).

\bibitem{Ariel-2015} M. B. Ayun, A. Schwarzbaum, S. Rosenberg, \emph{et al.},
\enquote{Photonic radio frequency phase-shift amplification by radio frequency interferometry,} Optics Letters \textbf{40}, 4863--4866 (2015).

\bibitem{Ariel-2018} M. B. Ayun, E. Liokumovitch, D. Gotliv, \emph{et al.},
\enquote{Phase-shift-amplified interferometry,} Optics Letters \textbf{43}, 2402--2405 (2018).

\bibitem{Ariel-2020} E. Liokumovitch and S. Sternklar, \enquote{Heterodyne phase-shift-amplified interferometry with improved shot-noise-limited sensitivity and immunity to RIN,} Optics Letters \textbf{45}, 1950--1953 (2020).

\bibitem{OTD} A. Esteban-Mart\'{i}n, J. Garc\'{i}a-Monreal, F. Silva, \emph{et al.}
\enquote{Interferometric measurement of complex-field changes in transient detection imaging,} Optics Express \textbf{28}, 28782--28791 (2020).


\bibitem{VIM} BIPM, IEC, IFCC, ILAC, ISO, IUPAC, IUPAP, and OIML, ``International vocabulary of metrology --- Basic and general concepts and associated terms (VIM),'' 3rd ed., Joint Committee for Guides in Metrology, JCGM200:2012, entry 4.12. URL: \url{https://www.bipm.org/documents/20126/2071204/JCGM_200_2012.pdf}.

\bibitem{Carre-1966} P. Carr\'{e}, \enquote{Installation et utilisation du comparateur photo\'{e}lectrique et interf\'{e}rentiel du Bureau International des Poids et Mesures,} Metrologia \textbf{2}, 13--23 (1966).

\bibitem{Heydemann-1981} P. L. M. Heydemann, \enquote{Determination and correction of quadrature fringe measurement errors in interferometers,} Applied Optics \textbf{20}, 3382--3384 (1981).

\bibitem{Fernandez-2013} M. R. Fern\'{a}ndez-Ruiz, M. Li, and J. Aza\~{n}a, \enquote{Time-domain holograms for generation and processing of temporal complex information by intensity-only modulation processes,} Optics Express \textbf{21}, 10314--10323 (2013).

\bibitem{Kolner-1994} B. H. Kolner, \enquote{Space-time duality and the theory of temporal imaging,} IEEE Journal of Quantum Electronics \textbf{30}, 1951--1963 (1994).

\bibitem{Cumming-1957} R. C. Cumming, \enquote{The serrodyne frequency translator,} Proceedings IRE \textbf{45}, 175--186 (1957).

\bibitem{Johnson-2010} D. M. S. Johnson, J. M. Hogan, S.-w. Chiow, \emph{et al.},
\enquote{Broadband optical serrodyne frequency shifting,} Optics Letters \textbf{35}, 745--747 (2010).

\bibitem{Takeda-1982} M. Takeda, H. Ina, and S. Kobayashi, \enquote{Fourier-transform method of fringe-pattern analysis for computer-based topography and interferometry,} Journal of the Optical Society of America \textbf{72}, 156--160 (1982).

\bibitem{Caves1981} C. M. Caves, \enquote{Quantum-mechanical noise in an interferometer,}
Physical Review D \textbf{23}, 1693--1708, (1981).

\bibitem{Haus2000} H. A. Haus, \emph{Electromagnetic Noise and Quantum Optical Measurements} (Springer, 2000), Chap.~8.

\bibitem{BachorRalph2019} H.-A. Bachor and T. C. Ralph, \emph{A Guide to Experiments in Quantum Optics}, 3rd ed. (Wiley-VCH, 2019), Chap.~8.

\bibitem{SalehTeich2019} B. E. A. Saleh and M. C. Teich, \emph{Fundamentals of Photonics}, 3rd ed. (Wiley 2019), Chaps.~13 and 25.

\bibitem{Harris-1978} F. J. Harris, \enquote{On the use of windows for harmonic analysis with the discrete Fourier transform,} Proceedings of the IEEE \textbf{66}, 51--83 (1978).

\bibitem{Kingston-book} R. H. Kingston, \emph{Detection of Optical and Infrared Radiation} (Springer, 1978), Chap.~3.
\bibitem{YuenChan1983} H. P. Yuen and V. W. S. Chan, \enquote{Noise in homodyne and heterodyne detection,} Optics Letters \textbf{8}, 177--179 (1983).

\bibitem{Boll-1979} S. Boll, \enquote{Suppression of acoustic noise in speech using spectral subtraction,} IEEE Transactions on Acoustics, Speech, and Signal Processing \textbf{27}, 113--120, (1979).

\bibitem{Baili-2007} G. Baili, M. Alouini, D. Dolfi, \emph{et al.},
\enquote{Shot-noise-limited operation of a monomode high-cavity-finesse semiconductor laser for microwave photonics applications,} Optics Letters \textbf{32}, 650--652 (2007).


\end{thebibliography}
\end{document}